\documentclass[11pt]{article}

\usepackage{mathtools}
\usepackage{amssymb}
\usepackage{dsfont}
\usepackage{slashed}
\usepackage{fullpage}
\usepackage[colorlinks=true,linkcolor=blue,citecolor=magenta,linktocpage=true]{hyperref}
\usepackage{titlesec}
\titleformat*{\section}{\normalsize\bfseries}
\titleformat*{\subsection}{\normalsize\bfseries}
\titleformat*{\subsubsection}{\normalsize\bfseries}
\usepackage[numbers,sort,compress]{natbib}
\makeatletter
\renewcommand{\@dotsep}{10000}

\def\be#1\ee{\begin{align}#1\end{align}}

\def\q{\qquad}
\def\f{\frac}

\def\lb{\big\lbrace}
\def\rb{\big\rbrace}

\def\ip{\lrcorner\,}
\def\ipp{\ip\!\!\!\ip}

\def\de{\mathrm{d}}

\def\C{\mathcal{C}}

\def\H{\mathcal{H}}

\def\Q{\mathcal{Q}}

\numberwithin{equation}{section}

\begin{document}

\title{\Large{\textbf{\sffamily Electromagnetic duality and central charge from first order formulation}}}
\author{\sffamily Marc Geiller$^1$, Puttarak Jai-akson$^{2,3}$, Abdulmajid Osumanu$^{4,5}$, Daniele Pranzetti$^{2,6}$}
\date{\small{\textit{
$^1$Univ Lyon, ENS de Lyon, Univ Claude Bernard Lyon 1,\\ CNRS, Laboratoire de Physique, UMR 5672, F-69342 Lyon, France\\
$^2$Perimeter Institute for Theoretical Physics,\\ 31 Caroline Street North, Waterloo, Ontario, Canada N2L 2Y5\\
$^3$Department of Physics and Astronomy, University of Waterloo,\\ 200 University Avenue West, Waterloo, Ontario, Canada, N2L 3G1\\
$^4$Department of Applied Mathematics, University of Waterloo,\\ 200 University Avenue West, Waterloo, Ontario, Canada, N2L 3G1\\
$^5$Quantum Leap Africa, AIMS Rwanda Centre, Sector Remera, KN 3 Kigali, Rwanda\\
$^6$Universit\`a degli Studi di Udine,
via Palladio 8,  I-33100 Udine, Italy\\}}}

\maketitle
\vspace{-0.8cm}
\begin{abstract}
In the context of the infrared triangle there have been recent discussions on the existence and the role of dual charges. We present a new viewpoint on dual magnetic charges in $p$-form theories, and argue that they can be inherited from the charges of a first order formulation as a topological BF theory with potential. This happens because, depending on the spacetime dimension and on the form degree, the so-called translational gauge symmetries of BF theory become reducible and therefore admit zero-modes. Although such zero-modes lead to trivial symmetries of the $p$-form theory, they are associated with non-trivial charges. These turn out to be precisely the dual magnetic charges. The centrally-extended current algebra of electric and magnetic charges in the $p$-form theory then descends naturally from that of BF theory. This is an effort towards finding an existence criterion for dual charges.
\end{abstract}

\thispagestyle{empty}
\setcounter{page}{1}
\bigskip
\hrule
\tableofcontents



\section{Introduction}

The infrared triangle \cite{Strominger:2017zoo} establishes a relationship, in massless theories, between soft theorems \cite{Bloch:1937pw,Low:1954kd,Gell-Mann:1954wra,Low:1958sn,Kazes:1959aa,Yennie:1961ad,Weinberg:1965nx,Burnett:1967km,White:2011yy,Cachazo:2014fwa,Campiglia:2015qka,Campiglia:2016hvg,Conde:2016csj,Panchenko:2017lkw}, asymptotic symmetries \cite{Bondi:1962px,Sachs:1962zza,Sachs:1962wk,Brown:1986nw,Barnich:2010eb,Strominger:2013jfa,Strominger:2013lka,Barnich:2013axa,He:2014cra,He:2014laa,Mohd:2014oja,He:2015zea,Kapec:2015ena,Gabai:2016kuf,Campiglia:2021bap}, and memory effects \cite{1974SvA....18...17Z,Braginsky:1986ia,1987Natur.327..123B,Christodoulou:1991cr,Wiseman:1991ss,Thorne:1992sdb,Blanchet:1992br,vanHaasteren:2009fy,Favata:2010zu,Bieri:2013hqa,Tolish:2014bka,Tolish:2014oda,Winicour:2014ska,Wang:2014zls,Strominger:2014pwa,Susskind:2015hpa,Pasterski:2015zua,Lasky:2016knh,Donnay:2018ckb}. Among countless other new developments, thinking about these topics has in particular shed new light on the notion of dualities. The quintessential example is electromagnetic (EM for short) duality, which exchanges the electric and magnetic fields in Maxwell's theory. While the classical realization of this duality has of course been known for a while \cite{Olive:1995sw,EMFigueroa,Aschieri:2008zz} (see also \cite{Agullo:2018iya,Agullo:2016lkj,Agullo:2014yqa,Agullo:2018nfv,Huang:2019cja}), it is only recently that a new picture has emerged, connecting in particular soft photon theorems to dual large gauge transformations and magnetic soft charges \cite{Strominger:2015bla,Campiglia:2016hvg,Laddha:2017vfh,Nande:2017dba,Bhattacharyya:2017obx,Hamada:2017bgi,Hijano:2020szl}. Interestingly, this has also fostered work on dual gravitational charges \cite{Godazgar:2018qpq,Godazgar:2018dvh,Godazgar:2019dkh,Godazgar:2020kqd,Godazgar:2020gqd,Kol:2019nkc,Kol:2020zth,Kol:2020vet,Oliveri:2020xls,Seraj:2021qja}, in spite of there being no known gravitational analogue of EM duality in the full theory \cite{Argurio:2009xr,Bunster:2012km,Bunster:2013tc,Hortner:2019iip,Snethlage:2021lsf} (see \cite{Geiller:2020edh,Geiller:2020okp} for a notion of dual charges in 3-dimensional triad gravity and \cite{FP} for a gravitational duality at null infinity). Dual charges corresponding to soft theorems have also been discussed in the case of the massless scalar \cite{Campiglia:2017dpg,Francia:2018jtb,Campiglia_2019,Henneaux_2019}.

In the context of Maxwell's theory, these developments have motivated the study of (asymptotic) magnetic charges. This was done recently in \cite{Hosseinzadeh:2018dkh} using the so-called duality-symmetric formulation \cite{Zwanziger:1968rs,Zwanziger:1970hk,Bliokh:2012zr,Cameron_2012,Cardona:2015woa}, and in \cite{Freidel:2018fsk} (see also \cite{Mathieu:2019lgi, Mathieu:2021qfs}) by introducing magnetic edge modes on an extended phase space. The remarkable result of these constructions is that the electric and magnetic charges satisfy a current algebra with non-vanishing central charge (see however \cite{Henneaux:2020nxi}). This appearance of Ka\v c--Moody current algebras has been discussed in \cite{Nande:2017dba} (see also \cite{Witten:1995gf,Freed:2006yc} in a different context).

In the present note we sketch a new (and more speculative) viewpoint on the magnetic charges and the centrally-extended EM charge algebra. This exploits the first order formulation of Maxwell's theory as a constrained topological BF theory. The idea behind this proposal is the observation that EM duality swaps Maxwell's field equations $\de{*F}=0$ and the Bianchi identity $\de F=0$. These are second order equations, and as their names indicate the first one is an equation of motion while the second one is an identity. This therefore suggests to study the first order formulation, where instead of a single second order equation of motion one has two first order equations \cite{Gaona:2006td}. The first order formulation of Yang--Mills theories can be obtained from topological BF theory \cite{Horowitz:1989ng} supplemented by a potential \cite{Escalante:2012tb,Escalante:2012zz}. It is known that BF theory admits two types of charges (which we could suggestively call electric and magnetic), arising from two independent gauge symmetries, and that these charges form a centrally-extended current algebra \cite{Balachandran:1992qg} (see also \cite{Geiller:2017xad,Geiller:2020edh} in the case of 3-dimensional gravity as a BF theory and \cite{Freidel:2016bxd, Freidel:2019ees} in the case of first order 4-dimensional gravity). The idea is then to argue that Maxwell's theory could inherit its magnetic charge from BF theory, as well as the corresponding centrally-extended EM charge algebra. The catch, however, is that the magnetic charges of BF arise because of the topological nature of the theory, and the existence of so-called ``translational'' gauge symmetries. In Maxwell's theory, which is evidently not topological, this symmetry is broken. Depending on the dimensionality of spacetime, the translational symmetries can however be reducible \cite{Henneaux:1992ig}, and therefore admit zero-modes. Here we argue that in the 4-dimensional case these zero-modes can be identified with the magnetic gauge parameter of Maxwell's theory. The magnetic charges are then seen as arising from the reducible part of the broken translational symmetries of a topological theory. For $p$-form theory\footnote{With our conventions $p$ denotes the degree of the field strength $F=\de A$ of the $(p-1)$-form $A$.} in a $d$-dimensional spacetime, this is possible as long as $d-p>1$, and in this case the translation zero-mode and the magnetic gauge parameter both have degree $d-p-2$. Consistently, this argument can also be applied to 3-dimensional Maxwell theory to show that it does not admit magnetic charges.

More generally, this proposal is motivated by the following natural question: \textit{In a given theory, specified by a Lagrangian and admitting asymptotic charges, how can we know if there are ``hidden'' dual charges?} In the gravitational case it has already been suggested that a complete description of the charges (i.e. including the dual ones) should rely on the first order formulation \cite{Godazgar:2018qpq,Godazgar:2018dvh,Godazgar:2019dkh,Godazgar:2020kqd,Godazgar:2020gqd,Oliveri:2020xls,Freidel:2020xyx,Freidel:2020svx}, using e.g. the first order Einstein--Cartan Lagrangian with topological terms such as the Holst term \cite{Holst:1995pc}. In electromagnetism it is necessary to resort to a BF theory with potential in order to have a first order formulation. Since 4-dimensional Maxwell theory does indeed satisfy $d-p>1$, it admits magnetic charges. Instead of introducing them by hand or with the the extended Lagrangians of \cite{Hosseinzadeh:2018dkh,Freidel:2018fsk}, we argue that they are present in the first order BF-type formulation provided we take into account the reducibility of the translations.

This note is organized as follows. In Section \ref{sec:2} we recall the study of the charges and charge algebra in the case of topological BF theories with Abelian gauge group. This includes the derivation of the electric and magnetic BF charges and of their centrally-extended charge algebra. For completeness and in order to describe dual scalar fields as well, we consider a BF theory of $p$-forms in $d$-dimensional spacetimes. In Section \ref{sec:3} we briefly introduce $p$-form theories. We then explain in Section \ref{sec:4} how $p$-form theories can be written in a first order formulation by adding a potential to the $p$-form BF theories. We apply this to 4-dimensional Maxwell theory and derive our observation concerning the origin of the magnetic charges as zero-modes of BF translations. We also apply this idea to 3-dimensional Maxwell theory, where it shows consistently that there are no magnetic charges. This is reinterpreted as the non-reducibility of the translations in 3-dimensional BF theory. Finally, we present some perspectives in Section \ref{sec:5}.

Our conventions and notations are as follows. We work on a $d$-dimensional Lorentzian manifold $M$ with boundary $\partial M$. It is foliated by Cauchy slices $\Sigma$ with $(d-2)$-dimensional boundary $\partial\Sigma$. This boundary has poles, which can be understood as Wilson line singularities, surrounded by $(d-3)$-dimensional ``circles'' $\C$ providing a regularization of such singularities. This geometrical setup is depicted on Figure \ref{space} below. As we work with differential forms in $d$ dimensions, we recall that the wedge product $P\wedge Q=(-1)^{pq}Q\wedge P$ is a map $\Omega^p(M)\times\Omega^q(M)\to\Omega^{p+q}(M)$, where $P$ is a differential $p$-form on $M$. With the exterior derivative $\de:\Omega^p(M)\to\Omega^{p+1}(M)$ we have the Leibniz rule $\de(P\wedge Q)=\de P\wedge Q+(-1)^pP\wedge\de Q$. Finally, using a metric $g_{\mu\nu}$ on $M$ we can define the Hodge duality map $*:\Omega^p(M)\to\Omega^{d-p}(M)$, whose only property we will need is that $**P=(-1)^{p(d-p)+1}P$. The variables used throughout this work and their corresponding form degree are summarized in table \ref{degree} below. For the covariant phase space, we use the convention $\delta\de=+\,\de\delta$, and use $\ipp$ to denote the inner product in field space between a field variation and a field space differential form.
\begin{figure}[h]
\begin{center}
\includegraphics[scale=0.25]{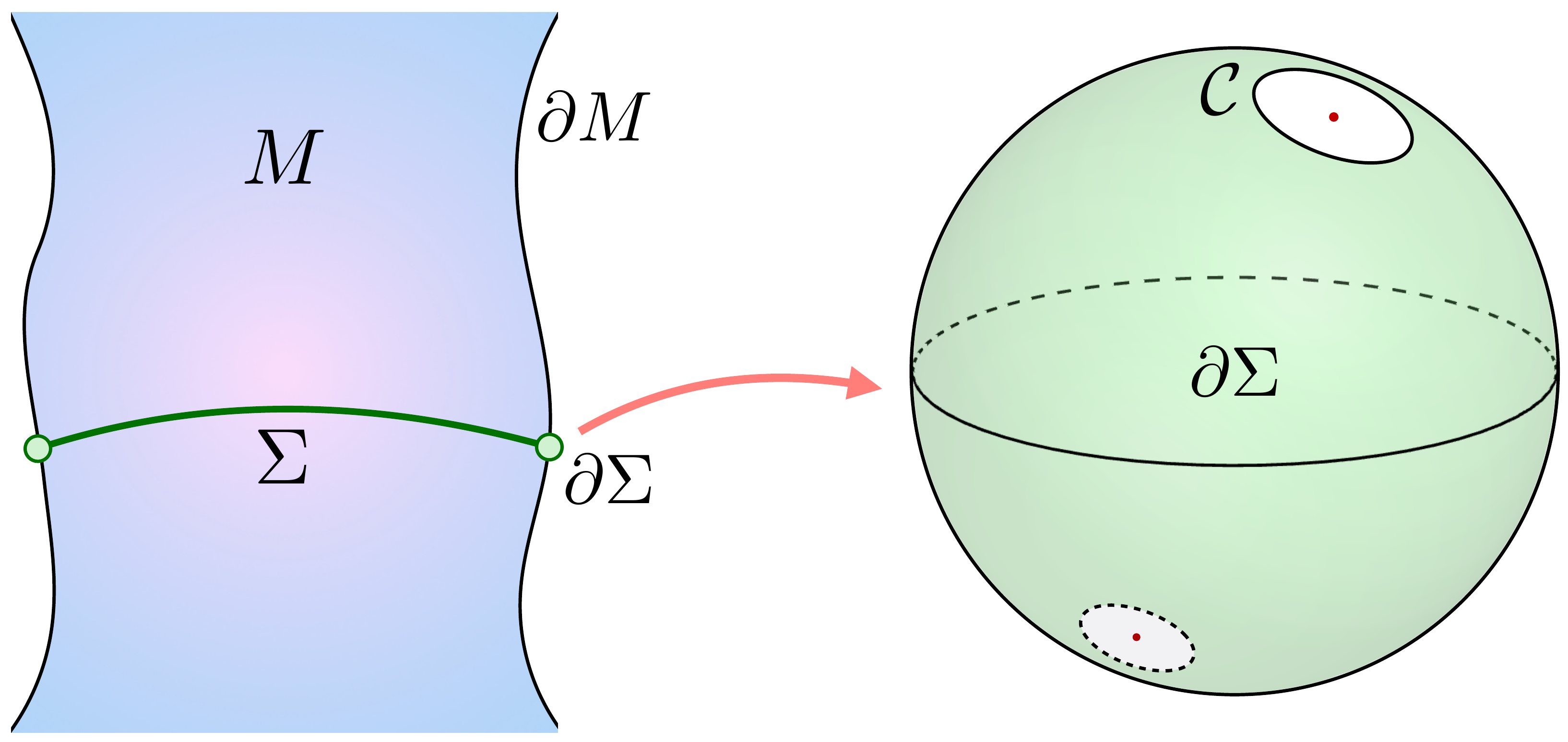}
\caption{A spacetime $M$ with boundary $\partial M$ and codimension-1 Cauchy slice $\Sigma$. The boundary of this latter is $\partial\Sigma=\Sigma\cap\partial M$. In the case where $\partial\Sigma$ has poles, we can think of $\partial\Sigma\backslash\{\text{poles}\}$ as having a set $\{\C\}$ of codimension-3 boundaries encircling these poles. In 4 dimension, this is the usual picture of a Dirac string piercing through the north and south poles of a 2-sphere.}\label{space}
\end{center}
\end{figure}

\begin{table}[h]
\begin{center}
\begin{tabular}{|c|c|c|c|c|c|c|}
\hline
variable & $\phantom{(\phi}A\phantom{\chi)}$ & $\phantom{(\phi}B\phantom{\chi)}$ & $\phantom{(}F=\de A\phantom{)}$ & $\phantom{F}(\alpha,\beta)\phantom{F}$ & $\phantom{F}(\phi,\chi)\phantom{F}$ & $\tilde{\alpha}$\\
\hline
form degree & $p-1$ & $d-p$ & $p$ & $p-2$ & $d-p-1$ & $d-p-2$\\
\hline
\end{tabular}
\caption{Variables and their associated form degree in $d$ dimension. Note that with our conventions when talking about a $p$-form theory the degree $p$ is that of the curvature $F=\de A$.}\label{degree}
\end{center}
\end{table}

\newpage

\section{BF theories}
\label{sec:2}

We begin by reviewing the covariant phase space of Abelian BF theory in the case of $p$-forms in $d$-dimensional spacetimes. This is a topological field theory whose basic fields are a connection $(p-1)$-form $A$ with curvature $F=\de A$, and a $(d-p)$-form $B$. The Lagrangian is
\be\label{BF Lagrangian}
L_\text{BF}[A,B]=F\wedge B,
\ee
where the order of wedge product has been chosen in order to minimize the amount of signs showing up below. Varying this Lagrangian gives
\be
\delta L_\text{BF}[A,B]=F\wedge\delta B+(-1)^p\delta A\wedge\de B+\de(\delta A\wedge B),
\ee 
which leads to the equations of motion
\be 
F=0,
\q\q
\de B=0.
\ee
The symplectic potential is $\theta=\delta A\wedge B$. Given a Cauchy slice $\Sigma\subseteq M$, the corresponding symplectic structure is
\be
\Omega_\Sigma=\int_\Sigma\delta\theta=-\int_\Sigma\delta A\wedge\delta B, 
\ee
and as usual it is independent of $\Sigma$ provided there is no symplectic flux leaking through the time-like boundary $\partial M$.

We now turn to the analysis of the Hamiltonian charges of BF theory. There are two kinds of conserved charges, associated with the two symmetries of the theory, namely the gauge symmetries and the translational symmetries (one can also use field-dependent combinations of these to describe diffeomorphisms).

\paragraph{Gauge symmetry.}

If $p\geq2$, the BF Lagrangian \eqref{BF Lagrangian} is invariant under the infinitesimal $\text{U}(1)$ gauge transformations
\be\label{gauge}
\delta^\text{(g)}_\alpha A=\de\alpha,
\q\q
\delta^\text{(g)}_\alpha B=0,
\ee
where $\alpha$ is a $(p-2)$-form. The Hamiltonian generator associated with this symmetry can be computed as
\be
\delta\H^\text{(g)}[\alpha]=-\delta^\text{(g)}_\alpha\ipp\Omega_\Sigma=\int_\Sigma\de\alpha\wedge\delta B.
\ee
On-shell, the gauge charges are therefore given by
\be\label{BF gauge charge}
\H^\text{(g)}[\alpha]=\int_\Sigma\de\alpha\wedge B\approx\int_{\partial\Sigma}\alpha\wedge B.
\ee  
These charges satisfy as expected a $\text{U}(1)$ current algebra
\be
\lb\H^\text{(g)}[\alpha],\H^\text{(g)}[\beta]\rb=-\delta^\text{(g)}_\alpha\ipp\delta^\text{(g)}_\beta\ipp\Omega_\Sigma=0.
\ee

\paragraph{Translational symmetry.}

By virtue of the Bianchi identity $\de F=0$, in the case $d-p\geq1$ the BF Lagrangian \eqref{BF Lagrangian} is also (quasi-)invariant under translational (or shift) symmetries, whose infinitesimal action is
\be\label{translation}
\delta^\text{(t)}_\phi A=0,
\q\q
\delta^\text{(t)}_\phi B=\de\phi,
\ee 
where $\phi$ is a $(d-p-1)$-form. The Hamiltonian generator is found from
\be
\delta \H^\text{(t)}[\phi]\coloneqq-\delta^\text{(t)}_\phi\ipp\Omega_\Sigma=-\int_\Sigma\delta A\wedge\de\phi,
\ee
from which we get that the translational charges are
\be\label{BF translation charge}
\H^\text{(t)}[\phi]=-\int_\Sigma A\wedge\de\phi\approx(-1)^p\int_{\partial \Sigma}A\wedge\phi.
\ee
These charges obey the Abelian algebra
\be
\lb\H^\text{(t)}[\phi],\H^\text{(t)}[\chi]\rb=-\delta^\text{(t)}_\phi\ipp\delta^\text{(t)}_\chi\ipp\Omega_\Sigma=0.
\ee

\paragraph{Central extension.}

The gauge and translational charges form a $\text{U}(1)\times\text{U}(1)$ Ka\v c--Moody algebra, where in addition to the brackets given above we have a central term given by the mixed bracket
\be\label{BF central bracket}
\lb\H^\text{(g)}[\alpha],\H^\text{(t)}[\phi]\rb=-\delta^\text{(g)}_\alpha\ipp\delta^\text{(t)}_\phi\ipp\Omega_\Sigma= (-1)^p\int_{\partial\Sigma}\de\alpha\wedge\phi.
\ee
Our goal is now to show that, when going from topological BF theory to a dynamical $p$-form theory (such as 4-dimensional Maxwell) the translational charge \eqref{BF translation charge} can survive as a magnetic charge, which then has a centrally-extended bracket \eqref{BF central bracket} with the electric charge.

\section{$\boldsymbol{p}$-form theories}
\label{sec:3}

We are interested in the EM duality for Abelian $p$-form theories, where $p$ is the degree of the curvature $F$. We note that asymptotic symmetries in $p$-forms theories were studied in \cite{Afshar:2018apx}. Let us first recall that the Lagrangian for such theories is
\be\label{p-form Lagrangian}
L_p[A]=\f{1}{2}{*F}\wedge F,
\ee 
where $F=\de A$ is again the curvature of the gauge field $A$. The equations of motion are
\be\label{p-form EOM}
\de{*F}=0,
\q\q
\de F=0,
\ee
where the second equation is the Bianchi identity. The theory is invariant under the action of $\text{U}(1)$ gauge transformations, whose finite form is $A\to A+\de\alpha$. The conserved charges associated with this gauge symmetry are the \textit{electric charges}
\be\label{E charge Maxwell}
\Q^\text{(E)}[\alpha]=\int_{\partial\Sigma}{*F}\wedge\alpha,
\ee
as can be worked out by computing the symplectic structure and contracting it with an infinitesimal gauge transformation.

The equations of motion \eqref{p-form EOM} and the Lagrangian suggest that interchanging $F$ and ${*F}$ leaves the theory unchanged. In other words, instead of using $F=\de A$ we can define ${*F}=\de\tilde{A}$ and work with $\tilde{A}$. This is the duality between a $p$-form and a $(d-p)$-form theory. In the case of Maxwell theory in 4 dimensions, which is a 2-form theory, this map is the \textit{EM duality}. In the general case of a $d$-dimensional $p$-form theory, this suggests that there must exist another type of charges, the \textit{magnetic charges}, of the form
\be\label{M charge}
\Q^\text{(M)}[\tilde{\alpha}]=\int_{\partial\Sigma}\tilde{\alpha}\wedge F.
\ee 
Since $\partial\Sigma$ is a codimension-2 manifold, a necessary condition for these magnetic charges to exist is that the form degree of the field strength $F$ be such that $p\leq(d-2)$, i.e. $d-p>1$. This ensures that $(d-p-2)$-forms exist, so that the wedge product of $F$ with a such a $(d-p-2)$-form $\tilde{\alpha}$ produces a $(d-2)$-dimensional form which can be integrated on the codimension-2 boundary $\partial\Sigma$. This is the reason for which magnetic charges cannot exist in e.g. 3-dimensional Maxwell theory.

As it is well-known,  differently than for the electric charge, the magnetic counterpart does not arise as the Noether charge of a bulk gauge transformation in the theory \eqref{p-form Lagrangian}. It is however possible to achieve this by changing the starting theory, and working instead with the so-called dual symmetric formulation as in \cite{Hosseinzadeh:2018dkh} or with an extended phase space as in \cite{Freidel:2018fsk}. Here we want to show that another understanding of these magnetic charges can be achieved from the first order formulation of the $p$-form theory, which we now present.

\section{First order $\boldsymbol{p}$-form theories from BF theories}
\label{sec:4}

The first order formulation of a $p$-form theory can be obtained as a BF theory with quadratic potential. The corresponding $d$-dimensional Lagrangian is
\be\label{first order Lagrangian}
L[A,B]=F\wedge B+\f{1}{2}{*B}\wedge B.
\ee
The presence of the metric in the definition of the Hodge dual breaks the topological nature of this theory. Its canonical analysis is performed in \cite{Escalante:2012tb,Escalante:2012zz}. To see that it indeed describes a $p$-form theory, we compte the variation
\be
\delta L[A,B]=(F+{*B})\wedge\delta B+(-1)^p\delta A\wedge\de B+\de(\delta A\wedge B),
\ee 
which gives the first order equations of motion
\be\label{BF+BB EOM}
F=-({*B})\ \Rightarrow\ B=(-1)^{p(d-p)}{*F},
\q\q
\de B=0.
\ee
Combining these leads to the second order $p$-form Maxwell equation $\de{*F}=0$. On-shell of the first equation of motion \eqref{BF+BB EOM}, the initial first order Lagrangian \eqref{first order Lagrangian} then reduces exactly to the $p$-form Lagrangian \eqref{p-form Lagrangian}.

Evidently, because of this on-shell equivalence, performing the analysis of the symmetries and of the charge algebra at the level of the first order $p$-form Lagrangian \eqref{first order Lagrangian} cannot a priori teach us anything valuable about the magnetic charges. This Lagrangian is indeed not invariant under the translational symmetry \eqref{translation} because of the presence of the potential term. Instead, our (admittedly non-standard) strategy will therefore be to first consider the charges and the charge algebra of BF theory alone, and only then impose in this structure the reduction to the non-topological $p$-form theory. The idea is simply to study how the reduction from BF theory to the $p$-form theory affects the charges discussed above in section \ref{sec:2}. The gauge symmetry $A\to A+\de\alpha$ survives this reduction, since this is still a symmetry of the $p$-form theory. Using the first equation of motion in \eqref{BF+BB EOM} shows that the BF gauge charge \eqref{BF gauge charge} becomes
\be\label{E charge}
\H^\text{(g)}[\alpha]\ \mapsto\ \Q^\text{(E)}[\alpha]=\int_{\partial\Sigma}{*F}\wedge\alpha,
\ee
which is the electric charge \eqref{E charge Maxwell}.

At first, the translational charges seem not to exist because, once the $B$ field is integrated out, the translation symmetry no longer survives. It is indeed evidently not a symmetry of the $p$-form theory. However, the subtlety is that the translational symmetry is reducible \cite{Henneaux:1992ig}, so that one should actually think of its breaking as a constraint on the transformation parameter $\phi$. This latter must be such that
\be\label{phi condition}
\de\phi=0\ \Rightarrow\ 
\left\{\begin{array}{lll}
\phi=\de\tilde{\alpha}&&\text{when }d-p>1\\
\phi=\text{const.}&&\text{when }d-p=1
\end{array}\right.
\ee
everywhere except at poles where $\de^2\tilde{\alpha}\neq0$. These restricted translations \textit{are} symmetries of the Lagrangian \eqref{first order Lagrangian}. When $d-p>1$, which is precisely the condition of existence of magnetic charges as explained below \eqref{M charge}, the reducible part of the translational symmetry is encoded in the $(d-p-2)$-form $\tilde{\alpha}$, which has the same form degree as the dual ${*\alpha}$ of the electric gauge parameter. With this identification, the translational charges become the magnetic charges as
\be\label{M charge 1}
\H^\text{(t)}[\phi]\ \mapsto\ \Q^\text{(M)}[\tilde{\alpha}]=(-1)^p\int_{\partial\Sigma}A\wedge\de\tilde{\alpha}.
\ee
Notice how this expression differs from the guess \eqref{M charge}. To understand this difference, we should recall that when poles are present the space $\partial\Sigma$ can be seen as having boundaries by cutting out all the poles. The resulting space, $\partial\Sigma\backslash\{\text{poles}\}$, is the $(d-2)$-dimensional space with small compact boundaries $\{ \C\}$ enclosing the poles, as in figure \ref{space}. This allows us to use integration by parts to obtain
\be \label{M charge 2}
\Q^\text{(M)}[\tilde{\alpha}]=\int_{\partial\Sigma\backslash{\{\text{poles}}\}}F\wedge\tilde{\alpha}-\sum_{\{\C\}}\oint_\C A\wedge\tilde{\alpha},
\ee
which agrees with the magnetic charge derived in \cite{Hosseinzadeh:2018dkh,Freidel:2018fsk}.

The algebra of electric and magnetic charges then inherits the central extension of BF theory, and in addition to the vanishing Abelian brackets we find that \eqref{BF central bracket} becomes
\be\label{central extension}
\lb\Q^\text{(E)}[\alpha],\Q^\text{(M)}[\tilde{\alpha}]\rb=(-1)^p\int_{\partial\Sigma}\de\alpha\wedge\de\tilde{\alpha}=-\sum_{\{\C\}}\oint_\C\de\alpha\wedge\tilde{\alpha},
\ee
which is also in agreement with \cite{Hosseinzadeh:2018dkh,Freidel:2018fsk}. It is now useful to study some explicit examples, such as Maxwell theory as well as scalar field theory and its dual.

\subsection{Maxwell theory}

For Maxwell theory the form degree is $p=2$. When $d=4$, we have the duality between the 2-forms $F$ and ${*F}$, and on the 2-sphere $\partial\Sigma=S^2$ we get the electric and magnetic charges
\be
\Q^\text{(E)}[\alpha]=\int_{S^2}\alpha\,{*F},
\q\q
\Q^\text{(M)}[\tilde{\alpha}]=\int_{S^2\backslash{\{\text{poles}}\}}\tilde{\alpha}F-\sum_{\{\C\}}\oint_\C\tilde{\alpha}A,
\ee
They form the $\text{U}(1)\times\text{U}(1)$ Ka\v c--Moody algebra
\be
\lb \Q^\text{(E)}[\alpha],\Q^\text{(M)}[\tilde{\phi}]\rb=-\sum_{\{\C\}}\oint_\C\tilde{\alpha}\de\alpha.
\ee
For other dimensions, the magnetic charges and the centrally-extended algebra exist in $p=2$ Maxwell theory as long as $d\geq4$, with the degree of the various fields given in table \ref{degree}.

For $d=3$ there is an electric charge, but since $\phi=\text{const.}$ according to \eqref{phi condition} there is only a global charge
\be
\Q^\text{(M)}_\text{global}[\phi]=\phi\int_{S^1}A,
\ee
and therefore no current algebra nor central extension.

Now, recall that while 4-dimensional Maxwell theory is self-dual, 3-dimensional Maxwell theory is dual to a scalar field theory. Let us therefore study the duality and first order formulation of a $d$-dimensional scalar field.

\subsection{Scalar field theory}

A free massless scalar field theory in $d$ dimensions can equivalently be thought of as a 1-form theory with Lagrangian 
\be\label{scalar Lagrangian}
L_\text{scalar}[\Phi]=\f{1}{2}{*\de\Phi}\wedge\de\Phi,
\ee
where the 0-form $\Phi\in\Omega^0(M)$ is the scalar field on $M$. Since this is not a gauge theory, it does not a priori admit conserved gauge charges. It is indeed obvious that the electric gauge transformations are not defined, since they require the form degree to be $p\geq2$. As such, this theory has no electric charges \eqref{E charge}.

It was however shown in \cite{Campiglia:2017dpg,Campiglia:2017xkp} that scalar field theories \textit{do} admit a reformulation of the soft theorem in terms of asymptotic symmetries, and contain \textit{scalar soft charges}. Following the argument built from the first order formulation, these are precisely the magnetic charges \eqref{M charge 1}, which in the scalar case become
\be
\Q^\text{(M)}[\tilde{\alpha}]=-\int_{\partial\Sigma}\Phi\de\tilde{\alpha},
\ee
where now $\tilde{\alpha}\in\Omega^{d-3}(M)$. This is consistent with the proposal of \cite{Francia:2018jtb,Campiglia_2019,Henneaux_2019}, which is to understand the scalar soft charges in terms of a gauge theory dual to the scalar field theory. In this reformulation, the above magnetic charges are interpreted as the electric charges of the gauge theory dual to the scalar theory \eqref{scalar Lagrangian}. This relies on the fact that a $d$-dimensional scalar field theory is dual to a gauge theory of $(d-1)$-forms via the identification
\be
F_{(d-1)} = \de A_{(d-2)} ={*\de\Phi}.
\ee
One can now apply our first order argument to this dual gauge theory, and verify by looking at the form degree that the electric charge exists while the magnetic one does not. The electric charge is given by
\be
\Q^\text{(E)}[\alpha]=\int_{\partial\Sigma}{*\big(F_{(d-1)}\big)}\wedge\alpha=(-1)^d\int_{\partial\Sigma}\de\Phi\wedge\alpha,
\ee
where $\alpha\in\Omega^{d-3}(M)$. We see that the magnetic charge in the scalar field theory agrees (up to an integration by parts and a possible sign) with the electric charge in the dual gauge theory. This charge, which is either magnetic for the scalar theory or electric for its dual, is the hidden scalar soft charge.

The same conclusion is reached when using the first order BF formulation. For a 4-dimensional scalar field, the dual is a theory of 3-forms. This latter has a first order formulation where the $B$ field is a 1-form. The translational symmetry is therefore not reducible, and no zero-mode survives the reduction from BF to the 3-form theory. Consistently, we get only a single charge for the scalar field or its dual, as described above.

\section{Perspectives}
\label{sec:5}

In this short note we have presented a new viewpoint on electromagnetic duality, motivated by recent studies of boundary magnetic charges and their centrally-extended algebra with electric charges \cite{Nande:2017dba,Hosseinzadeh:2018dkh,Freidel:2018fsk}. Instead of reformulating the second order Maxwell theory by introducing extra bulk or boundary fields as in \cite{Hosseinzadeh:2018dkh,Freidel:2018fsk}, we have followed the suggestion (which finds support in the gravitational case) that dual charges are naturally described in the first order formulation. We have described the first order formulation of Maxwell's theory as a topological BF theory with potential in Section \ref{sec:4}. In this first order formulation it is however still not clear where magnetic charges can come from. We have therefore put forward the following argument. A necessary condition for magnetic charges to exist in a $d$-dimensional $p$-form theory is that $d-p>1$. This condition turns out to be the one for which the translational symmetry of topological BF theory becomes reducible, and therefore admits zero-modes. These zero-modes have the same form degree as the would-be magnetic gauge parameters. Although the subset of translation symmetries corresponding to their reducible part describes trivial symmetries of the first order $p$-form Lagrangian \eqref{first order Lagrangian}, these symmetries can be assigned non-trivial charges which turn out to agree with the magnetic charges derived e.g. in \cite{Hosseinzadeh:2018dkh,Freidel:2018fsk}. Moreover, the electric and magnetic charges inherit the centrally-extended charge bracket descending from BF theory. In this picture the magnetic charges are the zero-modes of the translations of a topological theory.

To make this argument tighter one should study the treatment of reducible gauge transformations from the point of view of boundary charges. Our derivation of the magnetic charges does indeed rely on the expectation that a subset of the gauge charges can survive the imposition of the constraint reducing BF theory to Maxwell. It would also be interesting to study how this construction can be extended to the non-Abelian case. A new complication in this case is that the reducibility condition \eqref{phi condition} now involves a covariant derivative $\de_A\phi=0$, and cannot naively be solved without imposing a condition e.g. on the boundary field strength. Finally, we note that here we have focused on finite boundaries, and that it would be interesting to properly study the appearance of magnetic charges from BF theory asymptotically, following e.g. \cite{Henneaux:2020nxi} and \cite{Afshar:2018apx}.

\section*{Acknowledgements}
We would like to thank Laurent Freidel and Seyed Faroogh Moosavian for discussions and comments. We would also like to express our gratitude to Sylvain Carrozza, Bianca Dittrich, Ma{\"i}t{\'e} Dupuis, and Florian Girelli, for organizing the 2019 PI--Waterloo Quantum Gravity Winter Retreat at Camp Kintail, where this work was done. MG thanks Perimeter Institute for hospitality, and Blagoje Oblak for his enthusiasm when discussing the present idea. PJ's research is supported by the DPST Grant from the government of Thailand, and Perimeter Institute for Theoretical Physics. Research at Perimeter Institute is supported in part by the Government of Canada through the Department of Innovation, Science and Economic Development Canada and by the Province of Ontario through the Ministry of Colleges and Universities. This project has received funding from the European Union's Horizon 2020 research and innovation programme under the Marie Sklodowska-Curie grant agreement No. 841923.

\bibliography{Biblio.bib}
\bibliographystyle{Biblio}

\end{document}